\documentstyle[hip-artc,amsfonts,amssymb]{article} 
\volnumber{15}  \edyear{2002} \frompage{000} \topage{000}                
\recrevdate{17 October 2002}                                             

\def\*{\star}
\def\rightmark{Deformation Quantization, Superintegrability, and 
Nambu Mechanics} 
\def\leftmark{C. Zachos and T. Curtright} 
\title{Deformation Quantization, Superintegrability, and Nambu Mechanics} 
\authors{\twerm 
{Cosmas K Zachos$^{1}$ and Thomas L Curtright$^{2}$ %
\index{Zachos, C.}
\index{Curtright, T.}
}\\[2.812mm]
{\normalsize
\hspace*{-8pt}$^1$ High Energy Physics Division,Argonne National Laboratory, \\
Argonne, IL 60439-4815, USA\\[0.2ex] 
\hspace*{-8pt}$^2$ Department of Physics, University of Miami,\\
Box 248046, Coral Gables, Florida 33124, USA
}}

\abstract{
Phase Space is the framework best suited for quantizing superintegrable 
systems---systems with more conserved quantities than 
degrees of freedom. In this quantization method, the symmetry algebras 
of the hamiltonian invariants are preserved most naturally. 
We illustrate the power and simplicity of the method 
through new applications to nonlinear $\sigma$-models, 
specifically for Chiral Models and de Sitter 
$N$-spheres, where the symmetric quantum 
hamiltonians amount to compact and elegant expressions, 
in accord with the Groenewold-van Hove theorem.
Additional power and elegance is provided by the use of 
Nambu Brackets (linked to Dirac Brackets)
involving the extra invariants of 
superintegrable models.  
The quantization of Nambu Brackets is then successfully
compared to that of Moyal, validating Nambu's original proposal,
while invalidating other proposals.} 
 
\makeindex
\begin{document}
 
\maketitle
This is a pedagogical talk based on \cite{sphere}, with expanded context.
Highly symmetric quantum systems are often integrable, and, in 
special cases, superintegrable and exactly solvable \cite{winter}.
A superintegrable system of $N$ degrees of freedom has more than
$N$ independent invariants, and a maximally superintegrable one has $2N-1$ 
invariants. 
In the case of velocity-dependent potentials, when quantization 
of a classical system presents operator ordering ambiguities 
involving $x$ and $p$, the general consensus has long been 
\cite{velo} to select those orderings
in the quantum hamiltonian which maximally preserve the symmetries 
present in the corresponding classical hamiltonian. However, 
even for simple systems, such as $\sigma$-models considered here, 
such constructions may become involved and needlessly technical.

There is a quantization procedure ideally suited to this problem of selecting 
the quantum hamiltonian which maximally preserves integrability.
In contrast to conventional operator 
quantization, this problem  is addressed most cogently in Moyal's 
phase-space quantization formulation  \cite{moyal,cfz}, reviewed in 
\cite{czreview}. 
The reason is that the variables involved in it 
(``phase-space kernels" or ``Weyl-Wigner inverse transforms of operators") 
are c-number 
functions, like those of the classical phase-space theory, 
and have the same interpretation, although they involve $\hbar$-corrections
(``deformations"), in general---so $\hbar\rightarrow 0$ reduces them to the 
classical expressions.  It is only the detailed algebraic structure 
of their respective brackets and composition rules which contrast with 
those for the variables of the classical theory. This complete formulation is 
based on the 
Wigner Function (WF), which is a quasi-probability distribution function in 
phase-space, and comprises the kernel function of the density matrix.
Observables and transition amplitudes are  phase-space integrals of kernel
functions weighted by the WF, in analogy to statistical mechanics. Kernel
functions, however, unlike ordinary classical functions, multiply through the
$\*$-product, a noncommutative, associative, pseudodifferential operation, 
which encodes the entire quantum mechanical action and whose antisymmetrization 
(commutator) is the Moyal Bracket (MB) \cite{moyal,cfz,czreview}, the
quantum analog of the Poisson Bracket (PB). 

Groenewold's correspondence principle theorem \cite{groenewold} 
(to which van Hove's extension is often attached \cite{vanHove}) 
points out that, in general, there is no invertible linear map from all 
functions of phase space $f(x,p), g(x,p),...,$ to hermitean operators in 
Hilbert space ${\mathfrak Q}(f)$, ${\mathfrak Q}(g),...,$ such that
the PB structure is preserved, 
\begin{equation}
{\mathfrak Q} (\{ f,g\})= \frac{1}{i \hbar} ~
[ ~  {\mathfrak Q}(f)  , {\mathfrak Q} (g)~\Bigr ] ~.
\end{equation}
Instead, the Weyl correspondence map \cite{weyl,czreview} from functions to
ordered operators, ${\mathfrak W}(f) \equiv \frac{1}{(2\pi)^2}\int d\tau 
d\sigma dx dp ~f(x,p) \exp (i\tau ({ {\mathfrak  p}}-p)+i\sigma ({ 
{\mathfrak  x}}-x))$,
specifies a $\*$-product \cite{groenewold,czreview},  ~ 
${\mathfrak W} (f\star g) ={\mathfrak W} (f) ~   {\mathfrak W} (g)$,
and thus 
\begin{equation}
{\mathfrak W} (\{\! \{ f,g \}\!  \} )= \frac{1}{i \hbar} ~
\Bigl [ {\mathfrak W}(f)  , {\mathfrak W} (g) \Bigr ] ~.
\end{equation}   
The MB is defined as 
$\{\!\{ f, g \}\!\} \equiv (f \* g - g\* f)/ i\hbar$, 
and, as $\hbar\rightarrow 0$, ~~MB $\rightarrow$ PB. That is,
it is the MB instead of the PB which maps invertibly to the quantum commutator.
This relation underlies the foundation of phase-space quantization 
\cite{groenewold,czreview}. 

Conversely, given an arbitrary operator 
${\mathfrak F} ({\mathfrak x}, {\mathfrak p})$ 
consisting of operators ${\mathfrak x}$ and 
${\mathfrak p}$, one might imagine rearranging it by use of Heisenberg 
commutations to a canonical completely symmetrized 
(Weyl-ordered) form, in general with $O(\hbar )$ terms generated in the 
process. It might then be mapped uniquely to its Weyl-correspondent c-number 
kernel function $f$ in phase space ${\mathfrak x}  \mapsto x$,  and 
${\mathfrak p} \mapsto p$,  ~
${\mathfrak W}^{-1}( {\mathfrak F}) = f(x,p,\hbar)$. (In practice, there is 
a more direct inverse transform formula \cite{groenewold,czreview} which 
bypasses a need to rearrange to a canonical Weyl ordered form explicitly.)
Clearly, operators differing from each other by different orderings of 
their ${\mathfrak x}$s and ${\mathfrak p}$s correspond to kernel 
functions $f$ coinciding with each other at $O(\hbar^0)$, but different 
in $O(\hbar)$, in general.      Thus, a survey of all 
alternate operator orderings in a problem with such ambiguities amounts,
in deformation quantization, to a survey of the ``quantum correction" 
$O(\hbar)$ pieces of the respective kernel functions, ie the inverse 
Weyl transforms of those operators, and their study is greatly systematized 
and expedited. Choice-of-ordering problems then reduce to purely
$\*$-product algebraic ones, as the resulting preferred orderings are 
specified through particular deformations in the c-number kernel expressions 
resulting from the particular solution in phase space. 

In this phase-space quantization language, Hietarinta \cite{hietarinta} has 
investigated  the simplest integrable systems of velocity-dependent 
potentials. In each system, he has promoted the vanishing of the PB of the
classical invariant $I$ (conserved integral)  with the hamiltonian, 
$\{ H,I\}=0$, to the vanishing of its MB with the
hamiltonian, $\{\! \{ H_{qm} ,I_{qm} \}\!  \} =0$. 
This dictates quantum corrections, addressed
 perturbatively in $\hbar$: he has found $O(\hbar^2)$ corrections to the $I$s
and $H$, needed for quantum symmetry. 
The specification of the symmetric
hamiltonian then is complete, since 
 {\em the quantum hamiltonian in terms of
classical phase-space variables corresponds uniquely to the Weyl-ordered
expression for these variables in operator language}. 

We quantize nonlinear $N$-sphere $\sigma$-models and chiral models 
algebraically, to argue for the general principles of power and convenience 
in isometry-preserving quantization in phase space: The procedure of 
determining the proper symmetric quantum Hamiltonian yields remarkably 
compact and elegant expressions.
Briefly, we find \cite{sphere} that the maximal symmetry generator 
invariants are undeformed by 
quantization, but the Casimir invariants  of their MB algebras {\em are} 
deformed, in accord with the Groenewold-van Hove theorem.
Hence, the hamiltonians are also deformed, 
\begin{equation}
H_{qm}- H = {\hbar^2 \over 8}\Bigl ( \det g - 1 -N(N-1) \Bigr ) ,\label{sN}
\end{equation}
for $S^N$, where $\det g$ is the determinant of the hypersphere metric in 
the orthogonal projection, and the spectra are seen to be proportional to 
$l(l+N-1)$ for integer $l$. For $G\times G$ chiral models, we find,
for the corresponding group manifold quantities, 
\begin{equation}
H_{qm}- H = {\hbar^2 \over 8}\Bigl (\Gamma^b_{ac}~ g^{cd} \Gamma^a_{bd} -
f_{ijk}f_{ijk}\Bigr ) .  
\end{equation}

Quantization of maximally superintegrable systems in phase space has an 
unexpected application: it facilitates explicit testing of Nambu Bracket (NB)
\cite{nambu} quantization proposals, through direct comparison to the 
conventional quantum answers thus found. 

The classical evolution of all 
functions in phase space for such systems is alternatively specified through 
NBs \cite{sphere,hietnambu}, because the phase-space velocity 
is always perpendicular to the 
$2N$-dimensional phase-space gradients of $2N-1$ independent integrals of 
the motion, $L_i$. 
Thus, for an arbitrary phase-space function $k$ with no explicit time 
dependence, the classical evolution is fully specified by a phase-space 
jacobian which amounts to the Nambu Bracket:
\begin{eqnarray}
{dk\over dt}&=&  V {\partial (k,L_1,...,...,L_{2N-1}) \over \partial 
(q_1,p_{1},  q_2,p_{2},  ...,q_{N},p_{N})   } \nonumber    \\
&\equiv &V  \{k,L_1,...,L_{2N-1}\},
\end{eqnarray}
where the proportionality constant $V$ can be shown to be a time independent
function of the invariants.  E.g., on $S^2$, 
\begin{equation}
{dk\over dt}= {\partial (k,L_x,L_y,L_z ) \over \partial 
(x,p_x,y,p_y)   } ~,  \label{Rum}
\end{equation}
where the $L_i$s are invariants corresponding to translations
along two orthogonal meridians and circles of latitude, respectively.

However, consistent quantization of NBs has been considered problematic 
ever since their inception.
Nevertheless, Nambu's original quantization  prescription \cite{nambu} 
can, indeed, succeed, despite widespread expectations to the contrary.
Comparison to the standard Moyal deformation quantization 
vindicates \cite{sphere} Nambu's early quantization  prescription
(and invalidates other prescriptions), for systems such as $S^N$. E.g.,
for the above $S^2$, the MB result is actually equivalent to 
Nambu's quantum bracket \cite{sphere},
\begin{equation}
\frac{dk}{dt}   =\frac{-1}{2\hbar ^{2}}\Bigl [ k,L_{x},L_{y},L_{z}\Bigl ]_{\*}
=   \{\! \{ k, H_{qm} \}\!\} ~,
\end{equation}
where ~~$\left [ A, B \right ]_{\*}   \equiv i\hbar~ \{\!\{A,B \}\!\}$ and 
\begin{eqnarray}
\left[ A,B,C,D\right] _{\star } &\equiv &A\star \left[ B,C,D\right]
_{\star }-B\star \left[ C,D,A\right] _{\star }+C\star \left[
D,A,B\right] _{\star }-D\star \left[ A,B,C\right] _{\star }  \nonumber \\
&=&  [A,B]_\star \star [C,D]_\star + [A,C]_\star \star [D,B]_\star 
+ [A,D]_\star \star [B,C]_\star \nonumber \\ 
&\phantom{a}& +[C,D]_\star \star [A,B]_\star + [D,B]_\star \star [A,C]_\star 
+ [B,C]_\star \star [A,D]_\star~.   
\end{eqnarray}

It turns out that certain conditions often posited and not easy to 
satisfy in general in NB quantization schemes are either satisfied 
automatically for these superintegrable systems, or the quantization 
goes through consistently despite failing such unnecessary conditions. 
We thus stress the utility of phase space quantization as a comparison testing 
tool for NB quantization proposals. 
More elaborate isometries of general manifolds in such models are expected to 
yield to analysis similar to what has been illustrated for prototypes.
\section*{Acknowledgements}
We gratefully acknowledge helpful questions from R Sasaki, 
D Fairlie, Y Nutku, V Fleurov, J-P Dahl, and J Klauder. 
This work was supported in part by the US Department of Energy, 
Division of High Energy Physics, Contract W-31-109-ENG-38, and the NSF Award 
0073390. 
 
\vfill\eject
\end{document}